\documentclass[
preprint,
 amsmath,amssymb,
 aps, physrev,
]{revtex4-2}

\usepackage{graphicx}
\usepackage{dcolumn}
\usepackage{bm}

\usepackage{caption}
\usepackage{xcolor}
\usepackage{diagbox}
\usepackage{booktabs}
\usepackage{adjustbox}
\usepackage[mathlines]{lineno}
\usepackage{ragged2e}

\newcommand{\ketbratwoargs}[2]{|#1\rangle\langle#2|}
\newcommand{\E}{\mathbb{E}}
\newcommand{\R}{\mathbb{R}}
\newcommand{\Tr}{\text{tr}}

\begin{document}


\title{\textbf{Efficient sampling for Pauli-measurement-based shadow tomography in direct fidelity estimation} 
}%

\author{Hyunho Cha}
\email{Contact author: aiden132435@cml.snu.ac.kr}
\affiliation{NextQuantum and Department of Electrical and Computer Engineering, Seoul National University, Seoul 08826, Republic of Korea}
\author{Jungwoo Lee}
\email{Corresponding author: junglee@snu.ac.kr}
\affiliation{NextQuantum and Department of Electrical and Computer Engineering, Seoul National University, Seoul 08826, Republic of Korea}

\date{\today}

\begin{abstract}
A constant number of random Clifford measurements allows the classical shadow protocol to perform direct fidelity estimation (DFE) with high precision. However, estimating properties of an unknown quantum state is expected to be more feasible with random Pauli measurements than with random Clifford measurements in the near future. Inspired by the importance sampling technique applied to sampling Pauli measurements for DFE, we show that similar strategies can be derived from classical shadows. Specifically, we describe efficient methods using only local Pauli measurements to perform DFE with GHZ, W, and Dicke states, establishing tighter bounds (by factor of $14.22$ and $16$ for GHZ  and W/Dicke, respectively) on the number of measurements required for desired precision. These protocols are derived by adjusting the distribution of observables. Notably, they require no preprocessing steps other than the sampling algorithms.
\end{abstract}

\maketitle


\section{Introduction}
Advances in the preparation of many-body entangled quantum states have marked significant progress in the field of quantum information \cite{lacroix2020symmetry, roghani2018dissipative, reiter2016scalable, kraus2008preparation}. A critical step in these experiments is verifying that the system's state aligns with the desired one \cite{jozsa1994fidelity, wilde2013quantum, liang2019quantum, baldwin2023efficiently}. One application of this verification step is in the verifier-prover paradigm of cloud quantum computing \cite{morimae2013blind, takeuchi2018verification}. Here, an untrusted server (prover) prepares an entangled many-body state, and the client (verifier) must certify that the returned state matches the target. This is accomplished by the prover supplying multiple copies of the state, which the verifier then measures to validate the preparation. Traditional methods for assessing quantum state fidelity include quantum state tomography (QST), which reconstructs the full density matrix of a quantum state, offering detailed information \cite{smithey1993measurement, james2001measurement, thew2002qudit, haffner2005scalable, leibfried2005creation, lvovsky2009continuous, gross2010quantum, toth2010permutationally, shang2017superfast}. However, QST is resource-intensive and scales poorly with system size, posing significant challenges for large-scale quantum systems \cite{cramer2010efficient}. Recent approaches have introduced various techniques aimed at simplifying the processes of verification \cite{takeuchi2018verification, zhu2019efficient, dangniam2020optimal, huang2024certifying} and fidelity estimation \cite{flammia2011direct, zhang2021direct, seshadri2024versatile}. Among these, direct fidelity estimation (DFE) has gained attention for its potential to provide accurate fidelity measurements with fewer resources compared to full state tomography. Unlike full tomography, DFE focuses on directly estimating the fidelity without requiring a complete reconstruction of the quantum state, thus significantly reducing the measurement overhead. For some classes of states, efficient methods using Pauli measurement \cite{da2011practical, flammia2011direct, zhang2021direct, leone2023nonstabilizerness}, some relying on \textit{importance sampling} \cite{kahn1951estimation, kloek1978bayesian}, are known.

Meanwhile, the \textit{classical shadow} estimation technique is known for its pioneering approach to efficiently predicting numerous properties of unknown quantum states \cite{huang2020predicting}, and recent studies have begun to explore its variants \cite{akhtar2023scalable, hu2023classical}. Specifically, it can estimate any quantity related to a quantum state $\rho$, expressed as $\text{tr}(\rho O)$ for some linear operator $O$. The fidelity between $\rho$ and a particular pure state is in this form, which implies that classical shadows can be used for DFE.

A classical snapshot of $\rho$ is generated by performing a random unitary operation $U$ on 
$\rho$ followed by a computational basis measurement. The efficiency of classical shadow tomography is determined by the distribution from which $U$ is chosen. Theoretically, random Clifford measurements \cite{calderbank1998quantum, aaronson2004improved, patel2008optimal, nielsen2010quantum} are preferable in the sense that a constant number of measurements suffices to estimate the fidelity with \textit{any} pure target state. Detailed analyses of the performance of this measurement have been carried out in the context of DFE. Unfortunately, performing random Clifford measurements in a laboratory setting is currently impractical \cite{bharti2022noisy}. More feasible alternatives are random Pauli measurements \cite{nielsen2010quantum}. They are much more resource-efficient and need precise control over only single-qubit operations. The significant drawback of random Pauli measurements is that, in the worst case, exponentially many samples with respect to the locality of an observable are necessary for precise estimation \cite{huang2020predicting}. However, for observables associated with the GHZ state \cite{greenberger1989going} and the W/Dicke state \cite{dur2000three, cabello2002bell, stockton2003characterizing}, which have maximum locality, the number of measurements required is only polynomial in the number of qubits. These entangled states play a central role as computational resources in various quantum protocols \cite{d2004computational, ben2005fast, faist2020continuous, hayakawa2022quantum, berry2024analyzing}.

In this work, we show that for some classes of target states, random Pauli measurements can be efficiently synthesized with classical shadows for DFE. Our protocols are designed to improve previous methods (based on characteristic functions) in the nonadaptive local Pauli measurement setting \cite{flammia2011direct}. The central aspect is that we employ a target-specific distribution of Pauli measurements rather than the default uniform distribution. This is analogous to sampling measurements from the characteristic function $\chi_{\rho}(k)$, which depends on the target $\rho$. Accordingly, the shadow inversion step is adapted to align with the modified distribution of measurements. As a result, the bound on the number of measurements necessary to achieve a specified precision is tightened by a factor between 14.22 and 28.44, and between 16 and 32 for the GHZ state and the W/Dicke state, respectively. The proposed estimator is more {\it direct} than Ref. \cite{flammia2011direct} (as will be explained in the discussion), resulting in more accurate estimation with the same number of measurements. Moreover, it does not require any preprocessing steps or memory to store the distribution. The proposed method is analyzed in a more general setting in Section~\ref{section:general_real_observable}, and it can also be extended to other scenarios as in \cite{liu2019efficient, li2021verification}.

\section{DFE with importance sampling}
The fidelity between an unknown state and a target state can be estimated from few Pauli measurements using importance sampling techniques \cite{flammia2011direct}. It is carried out in two stages: first by sampling a Pauli observable, followed by estimating its corresponding characteristic function. The fidelity is estimated to an additive error of $2\epsilon$ with probability $\geq 1 - 2\delta$ using
\begin{equation*}
N \leq l + 1 + \frac{2 \log (2 / \delta)}{\epsilon^2} \alpha
\end{equation*}
measurements, where we may choose
\begin{equation}
\label{equation:estimation_params}
(l, \alpha) = \left(\frac{2\log (2 / \delta)}{\epsilon^2}, 1 \right) \quad \text{and} \quad (l, \alpha) = \left(\frac{1}{\epsilon^2 \delta}, n^2 \right)
\end{equation}
for the GHZ state
\begin{equation*}
|\text{GHZ}\rangle = \frac{|0\rangle^{\otimes n} + |1\rangle^{\otimes n}}{\sqrt{2}}
\end{equation*}
and the W state
\begin{equation*}
|\text{W}\rangle = \frac{|100\cdots 0\rangle + |010\cdots 0\rangle + \cdots + |00\cdots 01\rangle}{\sqrt{n}},
\end{equation*}
respectively. Replacing $(\epsilon, \delta)$ with $(\epsilon / 2, \delta / 2)$, we can estimate the fidelity to an additive error of $\epsilon$ with probability $\geq 1 - \delta$ using less than
\begin{equation}
\label{equation:num_samples_approx}
\frac{16 \log (4 / \delta)}{\epsilon^2} \quad \text{and} \quad \frac{8 \log(4 / \delta)}{\epsilon^2} n^2
\end{equation}
measurements for the GHZ state and the W state, respectively. The choices for $l$ in (\ref{equation:estimation_params}) and the approximations in (\ref{equation:num_samples_approx}) are given because asymptotics with respect to $(\epsilon, \delta)$ are considered for the GHZ state, whereas for the W state, asymptotics with respect to $n$ are considered while keeping $(\epsilon, \delta)$ constant.

\section{Classical shadows}
Many properties of a quantum state can be predicted without its full characterization. Such processes are referred to as \textit{shadow tomography}. Throughout this work, $\rho$ denotes an unknown quantum state in $d = 2^n$ dimensions, associated with an $n$-qubit system. Let $\mathcal{U}$ be a distribution over unitary operators in a $d$-dimensional Hilbert space. Repeating the following steps generates random classical shapshots of $\rho$:
\begin{enumerate}
\item Sample $U \sim \mathcal{U}$.
\item Apply $U$ to $\rho$ ($\rho \mapsto \rho' = U \rho U^\dagger$).
\item Measure $\rho'$ in the computational basis.
\item Upon receiving the measurement outcome $\hat{b} \in \{0, 1\}^n$, store an efficient description of $U^\dagger |\hat{b}\rangle\langle\hat{b}| U$ in classical memory.
\end{enumerate}
This snapshot, averaged over $U$ and $\hat{b}$, can be viewed as a quantum channel \cite{holbrook2003noiseless, nielsen2010quantum, weedbrook2012gaussian}
\begin{equation*}
\mathcal{E}(\rho) \equiv \mathbb{E}[U^\dagger |\hat{b}\rangle\langle\hat{b}| U],
\end{equation*}
where $\mathcal{E}$ depends on $\mathcal{U}$. If $\mathcal{U}$ defines a tomographically complete set of measurements, then $\mathcal{E}$ has an inverse and the snapshot of $\rho$ is defined as
\begin{equation*}
\hat{\rho} = \mathcal{E}^{-1}[U^\dagger |\hat{b}\rangle\langle\hat{b}| U],
\end{equation*}
where $\mathbb{E}[\hat{\rho}] = \rho$.

\begin{figure}[t]
\centering
\begin{minipage}[b]{0.9\textwidth}
\centering
\includegraphics[width=\textwidth]{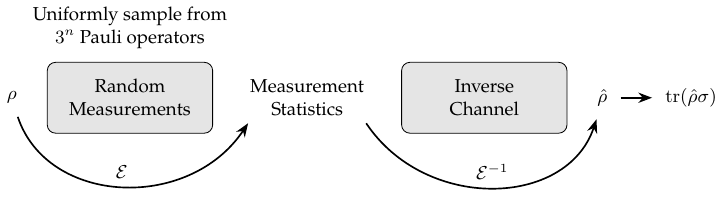}
(a)
\end{minipage}
\par\bigskip
\begin{minipage}[b]{0.8\textwidth}
\centering
\includegraphics[width=\textwidth]{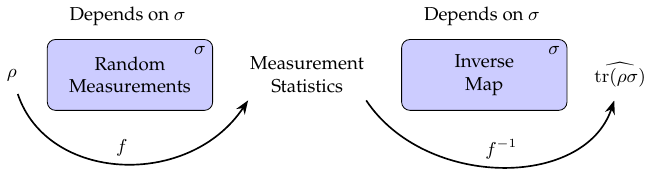}
(b)
\end{minipage}
\captionsetup{labelformat=empty, justification=raggedright, singlelinecheck=false}
\begin{nolinenumbers}
\caption{\label{figure:schematic}FIG. 1. (a) Vanilla implementation of DFE with classical shadows. (b) Sample-efficient implementation of DFE derived from classical shadows.}
\end{nolinenumbers}
\end{figure}

In many cases, a useful property of $\rho$ is given in the form of the expectation value of an observable:
\begin{equation}
\label{equation:expectation_observable}
\langle O \rangle = \text{tr}(\rho O).
\end{equation}
A snapshot of this quantity is $\text{tr}(\hat{\rho} O)$ and its average equals $\langle O \rangle$. One important application of this technique is in DFE. We restrict our attention to estimating the fidelity with a \textit{pure} target state
\begin{equation*}
\sigma = |\psi\rangle\langle\psi|.
\end{equation*}
The most straightforward way to estimate the fidelity
\begin{equation*}
F(\rho, \sigma) = \langle\psi|\rho|\psi\rangle = \text{tr}(\rho\sigma)
\end{equation*}
is to substitute $O$ with $\sigma$ for the snapshot, i.e., calculate the average of $\text{tr}(\hat{\rho} \sigma)$ as shown in FIG.~\ref{figure:schematic}(a). However, this approach can be highly inefficient, and our objective is to derive target-dependent modifications of this process as in FIG.~\ref{figure:schematic}(b).

\section{Random measurements}
Although classical shadows can be defined for any distribution $\mathcal{U}$, the following two distributions are mathematically convenient and widely discussed in literature:

\subsection{Random $n$-qubit Clifford circuits}
The $n$-qubit Clifford group $\text{Cl}(2^n)$ is a group of $n$-qubit unitary operators generated by the gates $\{H, S, \text{CNOT}\}$. A circuit $U$ is chosen uniformly from this group and the associated snapshot is
\begin{equation*}
\hat{\rho} = (2^n + 1)U^\dagger|\hat{b}\rangle\langle\hat{b}|U - \mathbb{I},
\end{equation*}
where $\mathbb{I}$ denotes the $2^n$-dimensional identity operator. A classical shadow of size independent of $n$ is sufficient to accurately estimate the fidelity with any pure target state \cite[Theorem 1]{huang2020predicting}.

\subsection{Tensor products of random single-qubit Clifford circuits}
In laboratory settings, random circuits in $\text{Cl}(2^n)$ are challenging to implement in the near future due to the presence of entangling gates such as $\text{CNOT}$. Practical circuits are those that consist only of local (single-qubit) unitaries. Throughout the rest of this paper, we focus exclusively on these circuits. For each qubit $i$, a single-qubit unitary $U_i$ is chosen uniformly and independently from $\text{Cl}(2)$. Instead, we can simply measure each qubit in a random Pauli basis chosen uniformly from
\begin{equation*}
S = \{X, Y, Z\},
\end{equation*}
which is equivalent to choosing $U_i$ uniformly from
\begin{equation*}
T = \{I, H, H S^\dagger\}.
\end{equation*}
In either case, the $n$-qubit unitary is $U = \bigotimes_{i=1}^{n} U_i$ and the associated snapshot is
\begin{equation}
\label{equation:snapshot}
\hat{\rho} = \bigotimes_{i=1}^{n}(3U_i^\dagger|\hat{b}_i\rangle\langle\hat{b}_i|U_i - \mathbb{I}).
\end{equation}
The worst case upper bound on the number of measurements to estimate (\ref{equation:expectation_observable}) scales exponentially with the locality of $O$ \cite[Proposition 3]{huang2020predicting} or the stabilizer R\'enyi entropy \cite[Theorem 1]{leone2023nonstabilizerness}. However, there exist much efficient methods for DFE with some well-conditioned states (as defined in Ref. \cite{flammia2011direct}), despite the locality of $O$ being $n$.

\section{Fidelity estimation from Pauli measurements}
The remainder of this work will be devoted to describing efficient methods to estimate the fidelity $F(\rho, \sigma)$ using random Pauli measurements and classical shadows, where $\sigma$ is either the GHZ state or the W state. For these states, classical shadows allow us to exclude unnecessary circuits by exploiting underlying symmetries.

\subsection{Fidelity with computational basis states}
We begin with the trivial case of estimating the fidelity between $\rho$ and a computational basis state, interpreted in the context of classical shadows. Suppose we want to use classical shadows to estimate $F(\rho, |b\rangle\langle b|)$, where $b \in \{0, 1\}^n$. A naive approach is to sample a Pauli string uniformly from $S^n$ for each measurement. The associated circuit $\bigotimes_{i=1}^{n} U_i$ is an element of $T^{\otimes n}$. From (\ref{equation:snapshot}), the estimator of $F(\rho, |b\rangle\langle b|)$ is calculated as
\begin{align}
\label{equation:estimator_product_expression}
\langle b|\hat{\rho}|b\rangle & = \langle b|\bigotimes_{i=1}^{n}(3U_i^\dagger|\hat{b}_i\rangle\langle\hat{b}_i|U_i - \mathbb{I})|b\rangle \notag \\
& = \prod_{i=1}^{n}(3\langle b_i|U_i^\dagger|\hat{b}_i\rangle\langle\hat{b}_i|U_i|b_i\rangle - \langle b_i|b_i\rangle) \notag \\
& = \prod_{i=1}^{n}(3|\langle\hat{b}_i|U_i|b_i\rangle|^2 - 1).
\end{align}
More generally,
\begin{equation}
\label{equation:estimator_product_expression_general}
\langle b_1|\hat{\rho}|b_2\rangle = \prod_{i=1}^{n}(3\langle b_{1,i}|U_i^\dagger|\hat{b}_i\rangle\langle\hat{b}_i|U_i|b_{2,i}\rangle - \delta_{b_{1,i}, b_{2,i}}).
\end{equation}
Note that $|\langle\hat{b}_i|U_i|b_i\rangle| = 1 / \sqrt{2}$ whenever $U_i \in T \setminus \{\mathbb{I}\}$, which can be verified with simple calculations. That is, if we measure the $i$th qubit in the $X$ or $Y$ basis, then $3|\langle\hat{b}_i|U_i|b_i\rangle|^2 - 1$ is always $1 / 2$ (regardless of $\hat{b}_i$). This allows us to obtain a sample for each of the $3^n$ Pauli measurements just from a single $ZZ\cdots Z$ measurement outcome $|\hat{b}\rangle$, because we can replace some of the $Z$ measurements with $X$ or $Y$ measurements for which the multiplier is $1 / 2$ with certainty. Let $m = |\{i | \hat{b}_i = b_i\}|$. Substituting $U_i$ with $\mathbb{I}$ for all $i$ in (\ref{equation:estimator_product_expression}) gives
\begin{equation*}
\langle b|\hat{\rho}|b\rangle = 2^m (-1)^{n - m}.
\end{equation*}
Suppose we replace $a$ out of the measurements for $\{i | \hat{b}_i = b_i\}$ and $b$ out of the measurements for $\{i | \hat{b}_i = \overline{b_i}\}$ with $X$ or $Y$. Using the same $|\hat{b}\rangle$ gives
\begin{equation*}
\langle b|\hat{\rho}|b\rangle = 2^{m - a} (-1)^{n - m - b} (1 / 2)^{a + b}.
\end{equation*}
Since we have $2^{a + b}$ choices for replacement, the average of all $3^n$ estimator values evaluates to
\begin{align*}
\text{avg} & = \frac{1}{3^n}\sum_{a=0}^{m}\sum_{b=0}^{n - m}\frac{\binom{m}{a}\binom{n - m}{b}2^{a + b}2^{m - a}(-1)^{n - m - b}}{2^{a + b}}\\
& = \frac{1}{3^n}\sum_{a=0}^{m}\binom{m}{a}2^{m - a}\sum_{b=0}^{n - m}\binom{n - m}{b}(-1)^{n - m - b}\\
& = \frac{1}{3^n}(1 + 2)^m (1 - 1)^{n - m} \quad (\text{define }0^0 = 1) \\
& = \delta_{\hat{b}, b}.
\end{align*}
As expected, estimating the fidelity with the computational basis state $|b\rangle$ just amounts to repeating a $ZZ \cdots Z$ measurement and counting the number of the measurement outcomes equal to $|b\rangle$. This interpretation will support our further analysis to derive efficient methods to estimate the fidelity with the GHZ state or the W state using Pauli measurements and classical shadows.

Throughout the derivation of our methods for the GHZ state and the W state, we will frequently refer to Table~\ref{table:values_summary}.

\begin{table}[b]
\begin{nolinenumbers}
\caption{\label{table:values_summary}Values of local traces for each measurement basis and measurement outcome.}
\end{nolinenumbers}
\begin{ruledtabular}
\begin{tabular}{cccc}
\textbf{Measurement basis}&
\( X \)&
\( Y \)&
\( Z \)\\
\colrule
$U_i$ & $H$ & $H S^\dagger$ & $\mathbb{I}$\\
$\langle 0|U_i^\dagger|0\rangle\langle 0|U_i|1\rangle$ & $1 / 2$ & $-i / 2$ & $0$\\
$\langle 0|U_i^\dagger|1\rangle\langle 1|U_i|1\rangle$ & $-1 / 2$ & $i / 2$ & $0$\\
$\langle 1|U_i^\dagger|0\rangle\langle 0|U_i|0\rangle$ & $1 / 2$ & $i / 2$ & $0$\\
$\langle 1|U_i^\dagger|1\rangle\langle 1|U_i|0\rangle$ & $-1 / 2$ & $-i / 2$ & $0$
\end{tabular}
\end{ruledtabular}
\end{table}

\subsection{Fidelity with GHZ state}
The density matrix of the GHZ state is
\begin{equation*}
\sigma = \frac{1}{2}(|0\rangle\langle 0|^{\otimes n} + |1\rangle\langle 1|^{\otimes n} + |0\rangle\langle 1|^{\otimes n} + |1\rangle\langle 0|^{\otimes n}).
\end{equation*}
We estimate the diagonal part
\begin{equation*}
d = \text{tr}\left(\rho\frac{1}{2}(|0\rangle\langle 0|^{\otimes n} + |1\rangle\langle 1|^{\otimes n})\right)
\end{equation*}
and the off-diagonal part
\begin{equation*}
o = \text{tr}\left(\rho\frac{1}{2}(|0\rangle\langle 1|^{\otimes n} + |1\rangle\langle 0|^{\otimes n})\right)
\end{equation*}
separately, where $F(\rho, \sigma) = d + o$.

To estimate $d$, we repeat a $ZZ\cdots Z$ measurement with the estimator defined as
\begin{equation*}
\hat{d} = \begin{cases}
1 / 2 & \text{if } \delta_{\hat{b}, \mathbf{0}} + \delta_{\hat{b}, \mathbf{1}} = 1,\\
0 & \text{otherwise.}
\end{cases}
\end{equation*}
Also, we can estimate $\text{tr}(\rho|0\rangle\langle 1|^{\otimes n})$ by choosing a random Pauli string from $\{X, Y, Z\}^n$ and applying (3):
\begin{align}
\label{equation:zero_one_term}
\text{tr}(\rho|0\rangle\langle 1|^{\otimes n}) & = \mathbb{E}_{(U_i), \hat{b}}[\langle \mathbf{1}|\hat{\rho}|\mathbf{0}\rangle] \notag \\
& = \sum_{(U_i) \in T^n}\frac{1}{3^n}\mathbb{E}_{\hat{b}}[\langle \mathbf{1}|\hat{\rho}|\mathbf{0}\rangle] \notag \\
& = \sum_{(U_i) \in T^n}\frac{1}{3^n}\mathbb{E}_{\hat{b}}\left[3^n\prod_{i=1}^{n}\langle 1|U_i^\dagger|\hat{b}_i\rangle\langle\hat{b}_i|U_i|0\rangle\right] \notag \\
& = \sum_{(U_i) \in T^n}\mathbb{E}_{\hat{b}}\left[\prod_{i=1}^{n}\langle 1|U_i^\dagger|\hat{b}_i\rangle\langle\hat{b}_i|U_i|0\rangle\right].
\end{align}
However, for any $i$, if we measure the $i$th qubit in the $Z$ basis, i.e., $U_i = \mathbb{I}$, then Table~\ref{table:values_summary} shows that the whole product in (\ref{equation:zero_one_term}) evaluates to zero. Therefore, it suffices to measure each qubit in the $X$ or $Y$ basis, and we replace (\ref{equation:zero_one_term}) with
\begin{equation}
\label{equation:zero_one_term_nontrivial}
\sum_{(U_i) \in (T \setminus \{\mathbb{I}\})^n}\mathbb{E}_{\hat{b}}\left[\prod_{i=1}^{n}\langle 1|U_i^\dagger|\hat{b}_i\rangle\langle\hat{b}_i|U_i|0\rangle\right].
\end{equation}
From Table~\ref{table:values_summary}, the term inside the square bracket in (\ref{equation:zero_one_term_nontrivial}) is either real if
\begin{equation}
\label{equation:even_y}
|\{i \mid U_i = H S^\dagger\}| \equiv 0 \pmod{2}
\end{equation}
or pure imaginary otherwise. In either case, it has an absolute value of $1 / 2^n$. Meanwhile, $\text{tr}(\rho|1\rangle\langle 0|^{\otimes n})$ is the complex conjugate of (\ref{equation:zero_one_term_nontrivial}): the sample
\begin{equation}
\label{equation:twice_sample}
V= \prod_{i=1}^{n}\langle 1|U_i^\dagger|\hat{b}_i\rangle\langle\hat{b}_i|U_i|0\rangle + \prod_{i=1}^{n}\langle 0|U_i^\dagger|\hat{b}_i\rangle\langle\hat{b}_i|U_i|1\rangle
\end{equation}
evaluates to $\pm 1 / 2^{n - 1}$ if (\ref{equation:even_y}) holds and zero otherwise. This symmetry allows us to further reduce the sample space of $(U_i)$ as
\begin{equation*}
o = \sum_{\substack{(U_i) \in (T \setminus \{\mathbb{I}\})^n, \\ |\{i \mid U_i = H S^\dagger\}| \equiv 0 \pmod{2}}} \mathbb{E}_{\hat{b}} \left[\frac{V}{2}\right].
\end{equation*}
Let $T_{\text{GHZ}}$ and $\mathcal{U}_{\text{GHZ}}$ denote this space and the uniform distribution over this space, respectively, where $|T_{\text{GHZ}}| = 2^{n - 1}$. It is not hard to show that
\begin{equation}
\frac{V}{2} = \frac{(-1)^{|\{i \mid U_i = H S^\dagger\}| / 2 + |\{i \mid \hat{b}_i = 1\}|}}{2^n}.
\end{equation}
\begin{table}[b]
\captionsetup{labelformat=empty}
\begin{nolinenumbers}
\caption{\label{table:algorithm_ghz_state}ALG. 1. DFE for $n$-qubit GHZ state.}
\end{nolinenumbers}
\begin{ruledtabular}
\begin{tabular}{l}
\textbf{Input:} Unknown state $\rho$, error parameters $(\epsilon, \delta)$\\
\textbf{Output:} Estimate of the fidelity $F(\rho, \sigma)$, where $\sigma$ is the $n$-qubit GHZ state\\
$N \leftarrow \left\lceil 2 \log (2 / \delta) / \epsilon^2 \right\rceil$\\
$\texttt{sum} \leftarrow 0$\\
\texttt{for} \underline{\hspace{0.5em}} \texttt{in range}($N$):\\
\quad \texttt{if} $1 / 3 > X \sim \mathcal{U}_{(0, 1)}$:\\
\quad \quad $\hat{b} = \texttt{measureComputationalBasis}(\rho)$\\
\quad \quad $\texttt{sum} \, \texttt{+=} \, 3\left(\delta_{\hat{b}, \mathbf{0}} + \delta_{\hat{b}, \mathbf{1}}\right) / 2 - 3 / 4$\\
\quad \texttt{else}\\
\quad \quad Sample $U_i \in \{H, H S^{\dagger}\}$ uniformly and independently for $1 \leq i \leq n - 1$.\\
\quad \quad Choose $U_n \in \{H, H S^{\dagger}\}$ such that $|\{i \mid U_i = H S^\dagger\}| \mod{2} = 0$.\\
\quad \quad $U \leftarrow \bigotimes_{i=1}^{n} U_i$\\
\quad \quad $\rho' \leftarrow U \rho U^{\dagger}$\\
\quad \quad $\hat{b} = \texttt{measureComputationalBasis}(\rho')$\\
\quad \quad $\texttt{sum} \, \texttt{+=} \, 3 \cdot (-1)^{|\{i \mid U_i = H S^\dagger\}| / 2 + |\{i \mid \hat{b}_i = 1\}|} / 4$\\
\quad \texttt{end if}\\
\texttt{end for}\\
\texttt{return} $\tilde{F} = \, \texttt{sum} / N + 1 / 4$
\end{tabular}
\end{ruledtabular}
\end{table}
Considering that each $(U_i)$ is chosen with probability $1 / 2^{n - 1}$, we may write
\begin{equation*}
o = \sum_{(U_i) \in T_{\text{GHZ}}} \mathbb{E}_{\hat{b}}\left[\frac{1}{2^{n - 1}}\hat{o}\right] = \mathbb{E}_{(U_i) \sim \mathcal{U}_{\text{GHZ}}, \hat{b}} \hat{o},
\end{equation*}
where $\hat{o} \equiv 2^{n - 2} V$. Therefore, we have $0 \leq \hat{d} \leq 1 / 2$ and $-1 / 2 \leq \hat{o} \leq 1 / 2$.

Finally, $\hat{F}$ is defined as a sample generated by sampling $3 (\hat{d} - 1 / 4)$ with probability $1 / 3$ and $3 \hat{o} / 2$ with probability $2 / 3$. Then $|\hat{F}| \leq 3 / 4$ and
\begin{align*}
\label{equation:hat_F_expectation}
\mathbb{E}[\hat{F}] + \frac{1}{4} & = \frac{1}{3} \mathbb{E}[3(\hat{d} - 1 / 4)] + \frac{2}{3} \mathbb{E}[3\hat{o} / 2] + \frac{1}{4}\\
& = d + o \\
& = F(\rho, \sigma).
\end{align*}
We generate
\begin{equation}
\label{equation:shadow_ghz_numsamples}
N = \left\lceil \frac{9 \log (2 / \delta)}{8\epsilon^2} \right\rceil
\end{equation}
samples of $\hat{F}$. From Hoeffding's inequality \cite{hoeffding1994probability, kahane1960proprietes},
\begin{equation*}
\label{equation:ghz_hat_f_estimate}
P\left(\left|\overline{F} + \frac{1}{4} - F\right| \geq \epsilon\right) \leq \delta.
\end{equation*}
The total number of measurements is about $9 \log (2 / \delta) / 8\epsilon^2$, which tightens the bound (\ref{equation:num_samples_approx}) by a factor of $128(1 + \log_{2 / \delta} 2) / 9$, a constant between $128 / 9$ and $256 / 9$ (closer to $128 / 9$ for small $\delta$). Alg.~1 provides a complete pseudocode for the GHZ state. It should be noted that the measurement settings for fidelity estimation in Alg.~1 also appear in some verification protocols \cite{pallister2018optimal, li2020optimal}, but in our context these settings are further associated with a fidelity estimator defined by the specific measurement basis and outcome.

\subsection{Fidelity with W state}
The density matrix of the W state is
\begin{equation*}
\sigma = \frac{1}{n}\left(\sum_{i = 1}^n |e_i\rangle\langle e_i| + \sum_{i \neq j} |e_i\rangle\langle e_j|\right),
\end{equation*}
where
\begin{equation*}
|e_i\rangle \equiv |\underbrace{00 \cdots 0}_{\times (i - 1)} 1 \underbrace{00 \cdots 0}_{\times (n - i)}\rangle.
\end{equation*}
We estimate the diagonal part
\begin{equation*}
d = \text{tr}\left(\rho\frac{1}{n}\sum_{i = 1}^n |e_i\rangle\langle e_i|\right)
\end{equation*}
and the off-diagonal part
\begin{equation*}
o = \text{tr}\left(\rho\frac{1}{n}\sum_{i \neq j} |e_i\rangle\langle e_j|\right)
\end{equation*}
separately, where $F(\rho, \sigma) = d + o$.

To estimate $d$, we repeat a $ZZ\cdots Z$ measurement with the estimator defined as
\begin{equation*}
\hat{d} = \begin{cases}
1 / n & \text{if } \sum_{i = 1}^n \delta_{\hat{b}, e_i} = 1,\\
0 & \text{otherwise.}
\end{cases}
\end{equation*}
The off-diagonal estimator \(\hat{o}\) can be found in Appendix~\ref{appendix:w_state_off_diagonal}. We have $0 \leq \hat{d} \leq 1 / n$ and $-(n - 1) / 2 \leq \hat{o} \leq (n - 1) / 2$.

\begin{table}[b]
\captionsetup{labelformat=empty}
\begin{nolinenumbers}
\caption{\label{table:algorithm_w_state}ALG. 2. DFE for $n$-qubit W state.}
\end{nolinenumbers}
\begin{ruledtabular}
\begin{tabular}{l}
\textbf{Input:} Unknown state $\rho$, error parameters $(\epsilon, \delta)$\\
\textbf{Output:} Estimate of the fidelity $F(\rho, \sigma)$, where $\sigma$ is the $n$-qubit W state\\
$N \leftarrow \left\lceil 2 \log (2 / \delta) (n - 1)^2 / \epsilon^2 \right\rceil$\\
$\texttt{sum} \leftarrow 0$\\
\texttt{for} \underline{\hspace{0.5em}} \texttt{in range}($N$):\\
\quad \texttt{if} $1 / (n^2 - n + 1) > X \sim \mathcal{U}_{(0, 1)}$:\\
\quad \quad $\hat{b} = \texttt{measureComputationalBasis}(\rho)$\\
\quad \quad $\texttt{sum} \, \texttt{+=} \, (n^2 - n + 1) \left(2 \sum_{i = 1}^n \delta_{\hat{b}, e_i} - 1\right) / 2n$\\
\quad \texttt{else}\\
\quad \quad Sample $\{i, j\} \sim f$ and $U' \sim g$.\\
\quad \quad $U_i,U_j \leftarrow U'$\\
\quad \quad $U_k \leftarrow \mathbb{I} \quad \forall k \notin \{i, j\}$\\
\quad \quad $U \leftarrow \bigotimes_{m=1}^{n} U_m$\\
\quad \quad $\rho' \leftarrow U \rho U^{\dagger}$\\
\quad \quad $\hat{b} = \texttt{measureComputationalBasis}(\rho')$\\
\quad \quad $\texttt{sum} \, \texttt{+=} \, (n^2 - n + 1) \left(2 \delta_{\hat{b}_i, \hat{b}_j} - 1 \right) \delta_{\hat{b}_{[n] \setminus \{i, j\}}, \mathbf{0}} / 2n$\\
\quad \texttt{end if}\\
\texttt{end for}\\
\texttt{return} $\tilde{F} = \, \texttt{sum} / N + 1 / 2n$
\end{tabular}
\end{ruledtabular}
\end{table}

Finally, $\hat{F}$ is defined as a sample generated by sampling $(n^2 - n + 1) (\hat{d} - 1 / 2n)$ with probability $1 / (n^2 - n + 1)$ and $(n^2 - n + 1) \hat{o} / n(n - 1)$ with probability $n(n - 1) / (n^2 - n + 1)$. Then $|\hat{F}| \leq (n^2 - n + 1) / 2n$ and

\begin{align*}
\label{equation:hat_F_expectation}
\mathbb{E}[\hat{F}] + \frac{1}{2n} & = \frac{\mathbb{E}[(n^2 - n + 1) (\hat{d} - 1 / 2n)]}{n^2 - n + 1} + \frac{n(n - 1)\mathbb{E}[(n^2 - n + 1) \hat{o} / n(n - 1)]}{n^2 - n + 1} + \frac{1}{2n}\\
& = d + o \\
& = F(\rho, \sigma).
\end{align*}
We generate
\begin{equation}
\label{equation:shadow_w_numsamples}
N = \left\lceil \frac{\log (2 / \delta) (n^2 - n + 1)^2}{2\epsilon^2 n^2} \right\rceil
\end{equation}
samples of $\hat{F}$. From Hoeffding's inequality,
\begin{equation*}
\label{equation:w_hat_f_estimate}
P\left(\left|\overline{F} + \frac{1}{2n} - F\right| \geq \epsilon\right) \leq \delta.
\end{equation*}
The total number of measurements is about $\log (2 / \delta) n^2 / 2\epsilon^2$, which tightens the bound (\ref{equation:num_samples_approx}) by a factor of $16(1 + \log_{2 / \delta} 2)$, a constant between $16$ and $32$ (closer to $16$ for small $\delta$). Alg.~2 provides a complete pseudocode for the W state. For relevant definitions, see Appendix~\ref{appendix:w_state_off_diagonal}.

\subsection{Fidelity with Dicke state}
The density matrix of the Dicke state with $k$ excitations is
\begin{equation*}
\sigma = \binom{n}{k}^{-1} \left(\sum_{|\mathbf{i}| = k} |\mathbf{i}\rangle\langle \mathbf{i}| + \sum_{\substack{|\mathbf{i}| = |\mathbf{j}| = k,\\ |\mathbf{i}| \neq |\mathbf{j}|}} |\mathbf{i}\rangle\langle \mathbf{j}|\right),
\end{equation*}
where $|\mathbf{i}|$ denotes the Hamming weight of $\mathbf{i} \in \{0, 1\}^n$. We estimate the diagonal part
\begin{equation*}
d = \text{tr}\left(\rho \binom{n}{k}^{-1} \sum_{|\mathbf{i}| = k} |\mathbf{i}\rangle\langle \mathbf{i}|\right)
\end{equation*}
and the off-diagonal part
\begin{equation*}
o = \text{tr}\left(\rho \binom{n}{k}^{-1} \sum_{\substack{|\mathbf{i}| = |\mathbf{j}| = k,\\ |\mathbf{i}| \neq |\mathbf{j}|}} |\mathbf{i}\rangle\langle \mathbf{j}| \right)
\end{equation*}
separately, where $F(\rho, \sigma) = d + o$.

To estimate $d$, we repeat a $ZZ\cdots Z$ measurement with the estimator defined as
\begin{equation*}
\hat{d} = \begin{cases}
\binom{n}{k}^{-1} & \text{if } \sum_{|\mathbf{i}| = k} \delta_{\hat{b}, \mathbf{i}} = 1,\\
0 & \text{otherwise.}
\end{cases}
\end{equation*}
We have $0 \leq \hat{d} \leq \binom{n}{k}^{-1}$ and $-\binom{n}{k}^{-1} \cdot c_l \leq \hat{o}_l \leq \binom{n}{k}^{-1} \cdot c_l$, where the definition of \(c_l\) and the off-diagonal estimator \(\hat{o}_l\) can be found in Appendix~\ref{appendix:dicke_state_off_diagonal}. Moreover, let
\begin{equation*}
S \equiv \frac{1}{2} + \sum_{l = \max (0, 2k - n)}^{k - 1} c_l.
\end{equation*}
Finally, $\hat{F}$ is defined as a sample generated by sampling $2S \left( \hat{d} - \frac{1}{2} \binom{n}{k}^{-1} \right)$ with probability $1 / 2S$ and $S \hat{o}_l / c_l$ with probability $c_l / S$. Then $|\hat{F}| \leq \binom{n}{k}^{-1} \cdot S$ and
\begin{align*}
\mathbb{E}[\hat{F}] + \frac{1}{2} \binom{n}{k}^{-1} & = \frac{2S \mathbb{E} \left[ \hat{d} - \frac{1}{2} \binom{n}{k}^{-1} \right]}{2S} + \sum_{l = \max (0, 2k - n)}^{k - 1} \frac{c_l \mathbb{E}[S \hat{o}_l / c_l]}{S} + \frac{1}{2} \binom{n}{k}^{-1}\\
& = d + \sum_{l = \max (0, 2k - n)}^{k - 1} o_l \\
& = d + o \\
& = F(\rho, \sigma).
\end{align*}
We generate
\begin{equation*}
N = \left\lceil \frac{2 \log (2 / \delta) S^2}{\epsilon^2 \binom{n}{k}^2} \right\rceil
\end{equation*}
samples of $\hat{F}$. From Hoeffding's inequality,
\begin{equation*}
\label{equation:w_hat_f_estimate}
P\left(\left|\overline{F} + \frac{1}{2} \binom{n}{k}^{-1} - F\right| \geq \epsilon\right) \leq \delta.
\end{equation*}
\begin{table}[b]
\captionsetup{labelformat=empty}
\begin{nolinenumbers}
\caption{\label{table:algorithm_dicke_state}ALG. 3. DFE for $n$-qubit Dicke state with $k$ excitations.}
\end{nolinenumbers}
\begin{ruledtabular}
\begin{tabular}{l}
\textbf{Input:} Unknown state $\rho$, error parameters $(\epsilon, \delta)$\\
\textbf{Output:} Estimate of the fidelity $F(\rho, \sigma)$, where $\sigma$ is the $n$-qubit Dicke state with $k$ excitations\\
Define $c_l$ as in (\ref{equation:c_l_definition}).\\
$N \leftarrow \left\lceil 2 \log (2 / \delta) S^2 / \epsilon^2 \binom{n}{k}^2 \right\rceil, \quad S \leftarrow \frac{1}{2} + \sum_{l = \max (0, 2k - n)}^{k - 1} c_l$\\
$\texttt{sum} \leftarrow 0$\\
\texttt{for} \underline{\hspace{0.5em}} \texttt{in range}($N$):\\
\quad Sample $l$ with probability $c_l / S$ for $\max (0, 2k - n) \leq l \leq k - 1$ and $1 / 2S$ for $l = k$\\

\quad \texttt{if} $l = k$:\\
\quad \quad $\hat{b} = \texttt{measureComputationalBasis}(\rho)$\\
\quad \quad $\texttt{sum} \, \texttt{+=} \, 2S \binom{n}{k}^{-1} \left( \sum_{|\mathbf{i}| = k} \delta_{\hat{b}, \mathbf{i}} - \frac{1}{2} \right)$ \\
\quad \texttt{else}\\
\quad \quad Sample $\{\mathbf{i}, \mathbf{j}\} \sim f_l$ and $U_{m \in \text{s}(\mathbf{i}) \Delta \text{s}(\mathbf{j}) - \max (\text{s}(\mathbf{i}) \Delta \text{s}(\mathbf{j}))}^\prime \sim g$ independently.\\
\quad \quad $U_{m \in \text{s}(\mathbf{i}) \Delta \text{s}(\mathbf{j}) - \max (\text{s}(\mathbf{i}) \Delta \text{s}(\mathbf{j}))} \leftarrow U_m^\prime, \quad U_{\max ( \text{s}(\mathbf{i}) \Delta \text{s}(\mathbf{j}) )} \leftarrow \mathcal{F} (\{ U_m^\prime \})$\\
\quad \quad $U_{m \notin \text{s}(\mathbf{i}) \Delta \text{s}(\mathbf{j})} \leftarrow \mathbb{I}$\\
\quad \quad $U \leftarrow \bigotimes_{m=1}^{n} U_m$\\

\quad \quad $\rho' \leftarrow U \rho U^{\dagger}$\\
\quad \quad $\hat{b} = \texttt{measureComputationalBasis}(\rho')$\\
\quad \quad $x \leftarrow |\{ m \in \text{s}(\mathbf{i}) \Delta \text{s}(\mathbf{j}) | U_m = H S^\dagger \}| / 2 + |\{ m \in \text{s}(\mathbf{i}) - \text{s}(\mathbf{j}) \mid (U_m, \hat{b}_m) \in (H, 1), (H S^\dagger, 0) \}|$ \\
\hspace{1.033cm} +  $|\{ m \in \text{s}(\mathbf{j}) - \text{s}(\mathbf{i}) \mid \hat{b}_m = 1 \}|$ \\
\quad \quad $\texttt{sum} \, \texttt{+=} \, (-1)^x S \cdot \delta_{\hat{b}_{m \in \text{s}(\mathbf{i}) \cap \text{s}(\mathbf{j})}, \mathbf{1}} \delta_{\hat{b}_{m \notin \text{s}(\mathbf{i}) \cup \text{s}(\mathbf{j})}, \mathbf{0}} / \binom{n}{k}$\\
\quad \texttt{end if}\\
\texttt{end for}\\
\texttt{return} $\tilde{F} = \, \texttt{sum} / N + 1 / 2 \binom{n}{k}$
\end{tabular}
\end{ruledtabular}
\end{table}
The total number of measurements is $O(n^{2k})$, where the asymptotics are with respect to $n$, while $k$, $\epsilon$, and $\delta$ remain constant. As a detailed comparison, it follows from \cite{flammia2011direct} that estimating the fidelity with an \(n\)-qubit Dicke state \(\sigma\) with \(k\) excitations to additive error \(\epsilon\) and failure probability \(\le \delta\) needs roughly
\begin{equation}
\label{equation:dicke_flammia_asymptotic}
\frac{8\log(4/\delta)}{\epsilon^2}\binom{n}{k}^2 \approx \frac{8\log(4/\delta)}{\epsilon^2}\frac{n^{2k}}{(k!)^2}
\end{equation}
Pauli measurements. This follows from the fact that for every Pauli string \(W\in\{I,X,Y,Z\}^{\otimes n}\),
\[
|\Tr(\sigma W)| = 0 \quad \text{or} \quad |\Tr(\sigma W)| \ge \binom{n}{k}^{-1}.
\]
Our protocol leverages Dicke state symmetries to reduce the sample space. For large \(n\) (with fixed \(k\)) we have
\[
S \approx c_0 = \binom{n}{2k}\binom{2k}{0}\binom{2k-1}{k-1} \approx \frac{n^{2k}}{2(k!)^2}.
\]
Consequently, the required number of samples becomes
\begin{equation}
\label{equation:dicke_shadow_samples}
\frac{2\log(2/\delta)}{\epsilon^2} \frac{S^2}{\binom{n}{k}^2} \approx \frac{2\log(2/\delta)}{\epsilon^2} \frac{n^{4k}}{4(k!)^4} \frac{(k!)^2}{n^{2k}} = \frac{\log(2/\delta)}{2\epsilon^2} \frac{n^{2k}}{(k!)^2}.
\end{equation}
Equation \eqref{equation:dicke_shadow_samples} improves upon \eqref{equation:dicke_flammia_asymptotic} by the factor \(16(1 + \log_{2 / \delta} 2)\), which recovers the W state gain when \(k=1\). Alg.~3 provides a complete pseudocode for the Dicke state with $k$ excitations. For relevant definitions, see Appendix~\ref{appendix:dicke_state_off_diagonal}.

\section{A generic optimization rule}
\label{section:general_real_observable}
The protocols developed for GHZ, W, and Dicke states can be subsumed under a more general optimization framework. Consider a real-valued observable
\[
O = \sum_{b_1, b_2 \in \{0,1\}^n} c_{b_1, b_2} \ketbratwoargs{b_1}{b_2}, \quad c_{b_1, b_2}\in\R.
\]
For a fixed pair \((b_1, b_2)\) define
\[
\text{eq}(b_1,b_2) = \{i \mid b_{1,i} = b_{2,i}\}, \quad \text{neq}(b_1,b_2) = \{i \mid b_{1,i} = 1 - b_{2,i}\}.
\]
To estimate the symmetric combination \(\langle b_1|\rho|b_2\rangle + \langle b_2|\rho|b_1\rangle\), it suffices to perform a Pauli measurement that meets the following local basis rule:
\begin{itemize}
    \item qubits in \(\text{eq}(b_1,b_2)\) are measured in the \(Z\) basis.
    \item qubits in \(\text{neq}(b_1,b_2)\) are measured in the \(X\) or \(Y\) basis, with even number of \(Y\) basis across \(\text{neq}(b_1,b_2)\).
\end{itemize}
Let \(\mathcal{P}^{(b_1,b_2)}\) be the set of Pauli strings obeying this rule. Because \(O\) has real coefficients (so \(c_{b_1, b_2} = c_{b_2, b_1}\) when \(b_1\neq b_2\)), the expectation value can be reconstructed solely from
\begin{equation}
\label{equation:nontrivial_pauli}
\bigcup_{\substack{b_1,b_2\in\{0,1\}^n,\\c_{b_1,b_2}\ne0}} \mathcal{P}^{(b_1,b_2)}.
\end{equation}
After restricting the sampling domain to \eqref{equation:nontrivial_pauli} we reweight and renormalize the estimators and the sampling probabilities (which is itself a nontrivial process that depends on both the set \eqref{equation:nontrivial_pauli} and the coefficients \(c_{b_1,b_2}\)), replacing the original shadow estimation using all \(3^n\) Pauli operators. An immediate corollary is that Pauli strings containing an odd number of \(Y\) are never required when \(O\in\R^{2^n\times2^n}\).

Suppose the number of nonzero off-diagonal, right triangular elements of \(O\) is \(r\). Then, according to the measurement rule described above, we obtain a very rough upper bound on the number of measurement settings required to estimate \(\Tr(\rho O)\), which is \(1+rd/2\). If \(r=\text{poly}(n)\), this already yields an exponential reduction compared to the total number of local Pauli measurement settings, which is \(3^n=\left(\frac{3}{2}\right)^n d\).

Whenever the target state factorizes as a tensor product of independent subsystems—such as a collection of Bell pairs used as essential resources in distributed quantum computing \cite{buhrman2003distributed, cuomo2020towards, caleffi2024distributed}—one incurs no loss of optimality by simply performing the optimal measurement protocol on each subsystem in parallel. Moreover, to estimate every subsystem fidelity to within an additive error \(\epsilon\) with confidence at least \(1-\delta\), the total number of measurements need only grow logarithmically with the number of subsystems.

In quantum machine learning, one must estimate the quantity
\[
\Tr(|\psi(\mathbf{x};\boldsymbol\theta)\rangle\langle\psi(\mathbf{x};\boldsymbol\theta)|O),
\]
where
\[
|\psi(\mathbf{x};\boldsymbol\theta)\rangle = \prod_{l=1}^L V^{(l)}(\boldsymbol\theta) U^{(l)}(\mathbf{x}) |0\rangle^{\otimes n}
\]
for parameterized unitaries \(V^{(l)}(\boldsymbol\theta)\) and data-encoding unitaries \(U^{(l)}(\mathbf{x})\). There is growing interest in applying shadow tomography to observable estimation in variational quantum neural networks \cite{huang2023post, jerbi2024shadows}. As our approach can target a wide range of observables, it naturally extends to such applications.


\section{Numerical simulations}
Using the previous importance sampling technique, the average number of samples falls strictly below the bound (\ref{equation:num_samples_approx}) for many configurations. Furthermore, the bound (\ref{equation:num_samples_approx}) itself may be quite rough, as it involves two layers of estimation: one for importance sampling and another for characteristic function estimation. Thus, it is essential to experimentally demonstrate the improvement of our method over the importance sampling approach.

In practice, the precision of estimates in Algs 1 and  2 is not directly controllable, but the number of samples can be specified. Therefore, we compare Ref. \cite{flammia2011direct} with Algs 1 and  2 as follows: First, Ref. \cite{flammia2011direct} with parameters $(\epsilon, \delta)$ is used to estimate the fidelity. Next, Algs 1 and  2 are executed using the same number of samples as Ref. \cite{flammia2011direct}. Finally, the mean squared error (MSE) of the estimates is evaluated as \(\E\left[\left(\tilde{F} - F(\rho,\sigma)\right)^2\right]\), where \(\tilde{F}\) is the output of our estimation protocol.

\begin{table}[h]
\captionsetup{labelformat=empty}
\begin{nolinenumbers}
\caption{\label{table:algorithm_generate_fidelity}ALG. 4. Generating a random state with a specified fidelity relative to a target state.}
\end{nolinenumbers}
\begin{ruledtabular}
\begin{tabular}{l}
\textbf{Input:} Target state $\sigma$, fidelity $f$\\
\textbf{Output:} Random state $\rho$ such that $F(\rho, \sigma) = f$\\
$\rho$ = \texttt{randomDensityMatrix}() \\
$P \leftarrow I - \sigma$ \\
$\rho \leftarrow P \rho P$ \\
$\rho \, \texttt{*=} \, (1 - f) / \text{tr}(\rho)$ \\
\texttt{return} $\rho + f \sigma$
\end{tabular}
\end{ruledtabular}
\end{table}

\clearpage

\begin{figure}[t]
\centering
\begin{minipage}[b]{0.8\textwidth}
\centering
\includegraphics[width=\textwidth]{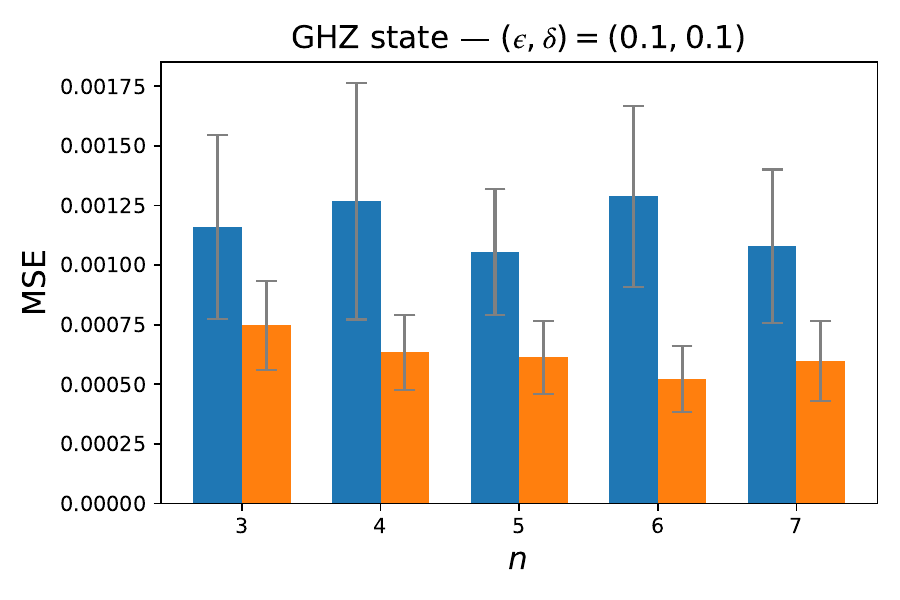}
(a)
\end{minipage}
\par\bigskip
\begin{minipage}[b]{0.8\textwidth}
\centering
\includegraphics[width=\textwidth]{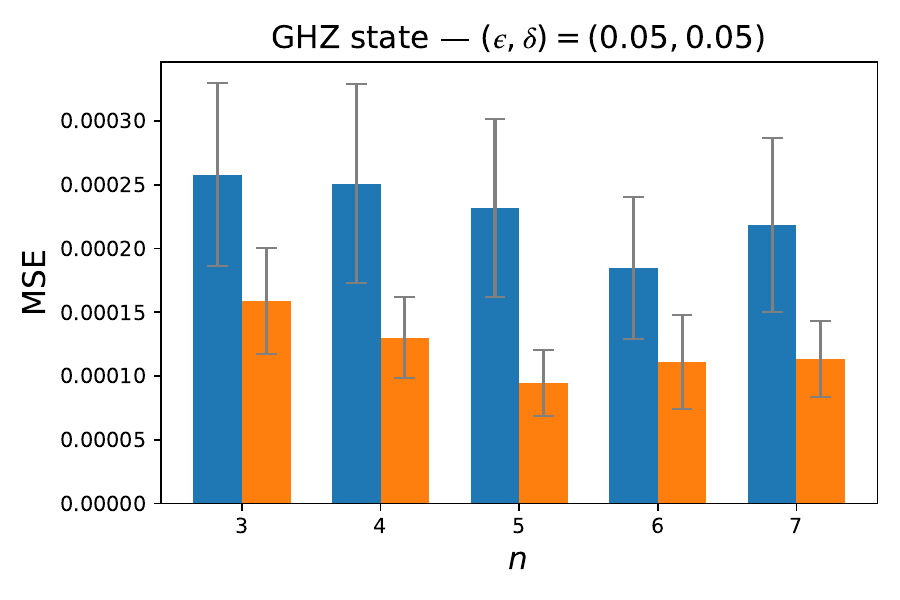}
(b)
\end{minipage}
\captionsetup{labelformat=empty, justification=raggedright, singlelinecheck=false}
\begin{nolinenumbers}
\end{nolinenumbers}
\end{figure}

\clearpage

\begin{figure}[t]
\centering
\begin{minipage}[b]{0.8\textwidth}
\centering
\includegraphics[width=\textwidth]{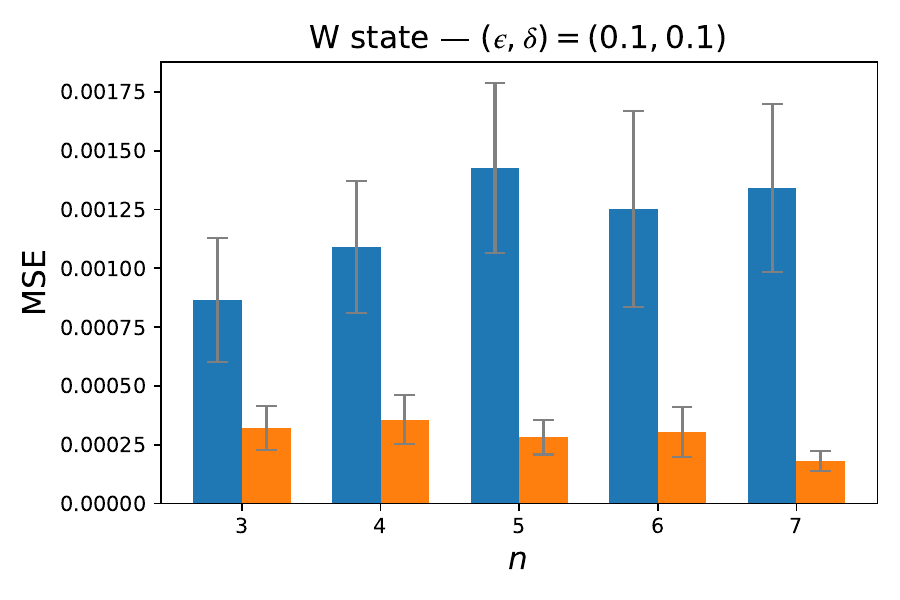}
(c)
\end{minipage}
\par\bigskip
\begin{minipage}[b]{0.8\textwidth}
\centering
\includegraphics[width=\textwidth]{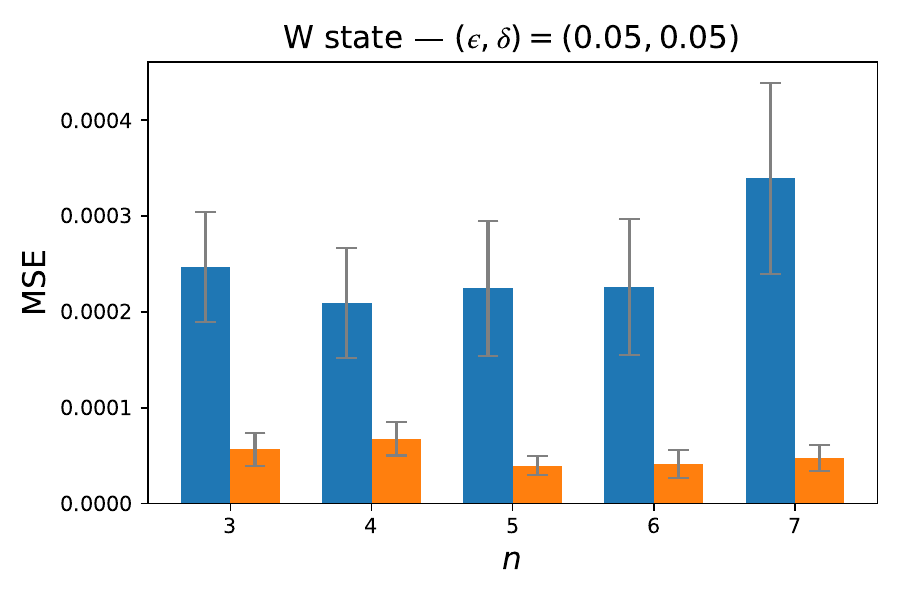}
(d)
\end{minipage}
\captionsetup{labelformat=empty, format=plain, singlelinecheck=false}
\begin{nolinenumbers}
\caption{\justifying\label{figure:plots}FIG. 2. MSE comparison for (a) GHZ state with \( (\epsilon, \delta) = (0.1, 0.1) \), (b) GHZ state with \( (\epsilon, \delta) = (0.05, 0.05) \), (c) W state with \( (\epsilon, \delta) = (0.1, 0.1) \), and (d) W state with \( (\epsilon, \delta) = (0.05, 0.05) \). Blue corresponds to the algorithm in Ref. \cite{flammia2011direct}, which is based on Pauli-Liouville coefficients \cite{chen2021robust}, and orange to our approach. Error bars represent 95\% confidence intervals.}
\end{nolinenumbers}
\end{figure}

\clearpage

The states were randomly generated with fidelities ranging from 0 to 1 in increments of 0.01 (i.e., $0, 0.01, 0.02, \ldots , 1$), as described in Alg. 4. Each state is generated by first creating a random mixed state and projecting it onto the subspace orthogonal to the target state (GHZ or W). The output state with the desired fidelity is then obtained as a linear combination of the target state and the projected state. FIG.~\ref{figure:plots} shows the comparison of MSE for up to 7 qubits. Algorithms 1 and  2 significantly improve estimation accuracy over Ref. \cite{flammia2011direct}, although the gain in this direction is less pronounced than the reduction implied by \eqref{equation:num_samples_approx} vs. \eqref{equation:shadow_ghz_numsamples} and \eqref{equation:shadow_w_numsamples}. The simulations were performed using PennyLane's classical simulation framework \cite{bergholm2018pennylane}.

\section{Discussion}
We have outlined efficient methods to estimate the fidelity between an unknown laboratory state and a pure target state using random Pauli measurement and classical shadows, where the target is either the GHZ state, the W state, or the Dicke state. The upper bounds on the required number of measurements to achieve a certain accuracy are tightened by constant factors. The improvement can be explained as follows.
\begin{itemize}
\item Instead of following the two-step procedure of sampling a Pauli operator and then estimating its characteristic function, we adopt a somewhat more direct way of estimating the fidelity based on a modified form of shadow tomography.
\item Local observables are sampled from $\{X, Y, Z\}^{\otimes n}$ rather than $\{\mathbb{I}, X, Y, Z\}^{\otimes n}$ (sampling the local observable $\mathbb{I}$ might be considered redundant since it always has an expectation value of 1). Moreover, deterministic and cancelled-out terms are carefully removed to further increase sample efficiency.
\end{itemize}
Optimizing the estimation of fidelities with states other than those considered in this work based on similar approaches is a promising direction for future research. Additionally, integrating recent techniques that employ strategies other than nonadaptive local Pauli measurements could yield further improvement.

Practical measurements suffer from noise, and the robust shadow estimation protocol was proposed to specifically mitigate this source of error \cite{chen2021robust}. Its calibration procedure for uniformly random Pauli measurements requires a number of samples that grows exponentially with the locality of the observable. Since every observable considered in this work is \(n\)-local, a direct application of that result would incur prohibitive overhead. However, our protocol departs from uniform Pauli sampling by tailoring the measurement ensemble to each target observable, and we believe that one can similarly devise bespoke calibration schemes whose sample complexity does not scale exponentially in \(n\). We therefore leave the development of such observable-specific calibration procedures as a key avenue for future work.

\begin{acknowledgments}
This work is in part supported by the National Research Foundation of Korea (NRF, RS-2024-00451435 (20\%), RS-2024-00413957 (40\%)), Institute of Information and  Communications Technology Planning and  Evaluation (IITP, 2021-0-01059 (40\%)), grant funded by the Ministry of Science and ICT (MSIT), Institute of New Media and Communications (INMAC), and the Brain Korea 21 FOUR program of the Education and Research Program for Future ICT Pioneers.
\end{acknowledgments}

\section*{Author Contributions}
H.C. conceptualized the study and conducted the experiments. J.L. supervised the project and provided feedback on the manuscript.

\appendix

\section{Off-diagonal terms for W state}
\label{appendix:w_state_off_diagonal}

For the off-diagonal part, consider the term
\begin{align}
\label{equation:one_hot_expression}
\text{tr}(\rho |e_i\rangle\langle e_j|) & = \mathbb{E}_{(U_k), \hat{b}}[\langle e_j|\hat{\rho}|e_i\rangle] \notag \\
& = \sum_{(U_k) \in T^n}\frac{1}{3^n}\mathbb{E}_{\hat{b}}[\langle e_j|\hat{\rho}|e_i\rangle] \notag \\
& = \sum_{\substack{U_i, U_j \in T \setminus \{\mathbb{I}\},\\ U_{k \notin \{i, j\}} = \mathbb{I}}}\frac{1}{3^2}\mathbb{E}_{\hat{b}}\Big[3^2 \langle 0|U_i^\dagger|\hat{b}_i\rangle\langle\hat{b}_i|U_i|1\rangle \langle 1|U_j^\dagger|\hat{b}_j\rangle\langle\hat{b}_j|U_j|0\rangle \delta_{\hat{b}_{[n] \setminus \{i, j\}}, \mathbf{0}}\Big] \notag \\
& = \sum_{\substack{U_i, U_j \in T \setminus \{\mathbb{I}\},\\ U_{k \notin \{i, j\}} = \mathbb{I}}}\mathbb{E}_{\hat{b}}\Big[\langle 0|U_i^\dagger|\hat{b}_i\rangle\langle\hat{b}_i|U_i|1\rangle \langle 1|U_j^\dagger|\hat{b}_j\rangle\langle\hat{b}_j|U_j|0\rangle \delta_{\hat{b}_{[n] \setminus \{i, j\}}, \mathbf{0}}\Big],
\end{align}
where $[n] \equiv \{1, 2, \cdots , n\}$. The third equality follows from previous analyses to reduce the sample space.

From Table~\ref{table:values_summary}, we see that the term inside the square bracket in (\ref{equation:one_hot_expression}) is zero if $\delta_{\hat{b}_{[n] \setminus \{i, j\}}, \mathbf{0}} = 0$, $\pm 1 / 4$ if $\delta_{\hat{b}_{[n] \setminus \{i, j\}}, \mathbf{0}} = 1$ and $U_i = U_j$, and $\pm i / 4$ if $\delta_{\hat{b}_{[n] \setminus \{i, j\}}, \mathbf{0}} = 1$ and $U_i \neq U_j$. Meanwhile, $\text{tr}(\rho |e_j\rangle\langle e_i|)$ is the complex conjugate of (\ref{equation:one_hot_expression}), and the sample
\begin{equation}
\label{equation:w_state_sample}
V_\text{W} = \left( \langle 0|U_i^\dagger|\hat{b}_i\rangle\langle\hat{b}_i|U_i|1\rangle \langle 1|U_j^\dagger|\hat{b}_j\rangle\langle\hat{b}_j|U_j|0\rangle + \langle 0|U_j^\dagger|\hat{b}_j\rangle\langle\hat{b}_j|U_j|1\rangle \langle 1|U_i^\dagger|\hat{b}_i\rangle\langle\hat{b}_i|U_i|0\rangle \right) \delta_{\hat{b}_{[n] \setminus \{i, j\}}, \mathbf{0}}
\end{equation}
evaluates to $\pm 1 / 2$ if $\delta_{\hat{b}_{[n] \setminus \{i, j\}}, \mathbf{0}} = 1$ and $U_i = U_j$. Otherwise, it is always zero. This symmetry allows us to further reduce the sample space of $(U_i)$:
\begin{equation*}
o = \sum_{i < j} \sum_{\substack{U_i = U_j \in T \setminus \{\mathbb{I}\},\\ U_{k \notin \{i, j\}} = \mathbb{I}}} \mathbb{E}_{\hat{b}} \left[\frac{V_\text{W}}{n}\right].
\end{equation*}
Let $f$ denote the uniform distribution over all $2$-subsets of $[n]$ and $g$ denote the uniform distribution over $T \setminus \{\mathbb{I}\}$. Then
\begin{align*}
o & = \sum_{i < j} \frac{2}{n(n - 1)} \sum_{U' \in T \setminus \{\mathbb{I}\}} \frac{1}{2} \mathbb{E}_{\hat{b}} \Big[\hat{o} \mid U_i = U_j = U', \, U_{k \notin \{i, j\}} = \mathbb{I}\Big]\\
& = \mathbb{E}_{\{i, j\} \sim f, \, U' \sim g, \hat{b}} \Big[\hat{o} \mid U_i = U_j = U', \, U_{k \notin \{i, j\}} = \mathbb{I}\Big],
\end{align*}
where $\hat{o} \equiv  (n - 1) V_\text{W}$. Therefore, we have $-(n - 1) / 2 \leq \hat{o} \leq (n - 1) / 2$.

\section{Off-diagonal terms for Dicke state}
\label{appendix:dicke_state_off_diagonal}

For the off-diagonal part, consider $\mathbf{i}, \mathbf{j} \in \{0, 1\}^n$ such that $|\mathbf{i}| = |\mathbf{j}| = k$ and $|\text{s}(\mathbf{i}) \cap \text{s}(\mathbf{j})| = l$, where $\text{s}(\mathbf{i}) \equiv \{ m | \mathbf{i}_m = 1 \}$ denotes the support of $\mathbf{i}$ and $l \in [\max (0, 2k - n), k - 1]$. For example, if
\begin{align*}
\mathbf{i} & = \begin{pmatrix} 1 & 1 & 1 & 1 & 1 & 0 & 0 & 0 & 0 & 0 \end{pmatrix}, \\
\mathbf{j} & = \begin{pmatrix} 0 & 0 & 0 & 1 & 1 & 1 & 1 & 1 & 0 & 0 \end{pmatrix},
\end{align*}
we have $n = 10$, $k = 5$, and $l = 2$. Then
\begin{align}
\label{equation:dicke_off_diagonal}
\text{tr}(\rho | \mathbf{i} \rangle\langle \mathbf{j} |) & = \mathbb{E}_{(U_m), \hat{b}}[\langle \mathbf{j} |\hat{\rho}| \mathbf{i} \rangle] \notag \\
& = \sum_{(U_m) \in T^n}\frac{1}{3^n}\mathbb{E}_{\hat{b}}[\langle \mathbf{j} |\hat{\rho}| \mathbf{i} \rangle] \notag \\
& = \sum_{\substack{U_{m \in \text{s}(\mathbf{i}) \Delta \text{s}(\mathbf{j})} \in T \setminus \{\mathbb{I}\},\\ U_{m \notin \text{s}(\mathbf{i}) \Delta \text{s}(\mathbf{j})} = \mathbb{I}}}\frac{1}{3^{2k - 2l}}\mathbb{E}_{\hat{b}}\left[3^{2k - 2l} \prod_{m \in \text{s}(\mathbf{i}) - \text{s}(\mathbf{j})} \langle 0|U_m^\dagger|\hat{b}_m\rangle\langle\hat{b}_m|U_m|1\rangle \right. \notag \\
& \hspace{5cm} \left. \prod_{m \in \text{s}(\mathbf{j}) - \text{s}(\mathbf{i})} \langle 1|U_m^\dagger|\hat{b}_m\rangle\langle\hat{b}_m|U_m|0\rangle \delta_{\hat{b}_{m \in \text{s}(\mathbf{i}) \cap \text{s}(\mathbf{j})}, \mathbf{1}} \delta_{\hat{b}_{m \notin \text{s}(\mathbf{i}) \cup \text{s}(\mathbf{j})}, \mathbf{0}} \right] \notag \\
& = \sum_{\substack{U_{m \in \text{s}(\mathbf{i}) \Delta \text{s}(\mathbf{j})} \in T \setminus \{\mathbb{I}\},\\ U_{m \notin \text{s}(\mathbf{i}) \Delta \text{s}(\mathbf{j})} = \mathbb{I}}} \mathbb{E}_{\hat{b}}\left[ \prod_{m \in \text{s}(\mathbf{i}) - \text{s}(\mathbf{j})} \langle 0|U_m^\dagger|\hat{b}_m\rangle\langle\hat{b}_m|U_m|1\rangle \prod_{m \in \text{s}(\mathbf{j}) - \text{s}(\mathbf{i})} \langle 1|U_m^\dagger|\hat{b}_m\rangle\langle\hat{b}_m|U_m|0\rangle \right. \notag \\
& \hspace{4cm} \left. \vphantom{\prod_{m \in \text{s}(\mathbf{j}) - \text{s}(\mathbf{i})}} \delta_{\hat{b}_{m \in \text{s}(\mathbf{i}) \cap \text{s}(\mathbf{j})}, \mathbf{1}} \delta_{\hat{b}_{m \notin \text{s}(\mathbf{i}) \cup \text{s}(\mathbf{j})}, \mathbf{0}} \right],
\end{align}
where $A \Delta B$ denotes the symmetric difference of the sets $A$ and $B$. The third equality follows from previous analyses to reduce the sample space.

From Table~\ref{table:values_summary}, we see that the term inside the square bracket in (\ref{equation:dicke_off_diagonal}) is zero if $\delta_{\hat{b}_{m \in \text{s}(\mathbf{i}) \cap \text{s}(\mathbf{j})}, \mathbf{1}} = 0$ or $\delta_{\hat{b}_{m \notin \text{s}(\mathbf{i}) \cup \text{s}(\mathbf{j})}, \mathbf{0}} = 0$, $\pm 1 / 2^{2k - 2l}$ if $\delta_{\hat{b}_{m \in \text{s}(\mathbf{i}) \cap \text{s}(\mathbf{j})}, \mathbf{1}} \delta_{\hat{b}_{m \notin \text{s}(\mathbf{i}) \cup \text{s}(\mathbf{j})}, \mathbf{0}} = 1$ and $|\{ m \in \text{s}(\mathbf{i}) \Delta \text{s}(\mathbf{j}) | U_m = H S^\dagger \}| \equiv 0 \pmod{2}$, and $\pm i / 2^{2k - 2l}$ if $\delta_{\hat{b}_{m \in \text{s}(\mathbf{i}) \cap \text{s}(\mathbf{j})}, \mathbf{1}} \delta_{\hat{b}_{m \notin \text{s}(\mathbf{i}) \cup \text{s}(\mathbf{j})}, \mathbf{0}} = 1$ and $|\{ m \in \text{s}(\mathbf{i}) \Delta \text{s}(\mathbf{j}) | U_m = H S^\dagger \}| \equiv 1 \pmod{2}$. Meanwhile, $\text{tr}(\rho | \mathbf{j} \rangle\langle \mathbf{i} |)$ is the complex conjugate of (\ref{equation:dicke_off_diagonal}), and the sample
\begin{align}
\label{equation:STISB_Dicke}
& V_\text{D} = \left( \prod_{m \in \text{s}(\mathbf{i}) - \text{s}(\mathbf{j})} \langle 0|U_m^\dagger|\hat{b}_m\rangle\langle\hat{b}_m|U_m|1\rangle \prod_{m \in \text{s}(\mathbf{j}) - \text{s}(\mathbf{i})} \langle 1|U_m^\dagger|\hat{b}_m\rangle\langle\hat{b}_m|U_m|0\rangle \right. \notag \\
& \left. + \prod_{m \in \text{s}(\mathbf{j}) - \text{s}(\mathbf{i})} \langle 0|U_m^\dagger|\hat{b}_m\rangle\langle\hat{b}_m|U_m|1\rangle \prod_{m \in \text{s}(\mathbf{i}) - \text{s}(\mathbf{j})} \langle 1|U_m^\dagger|\hat{b}_m\rangle\langle\hat{b}_m|U_m|0\rangle \right) \delta_{\hat{b}_{m \in \text{s}(\mathbf{i}) \cap \text{s}(\mathbf{j})}, \mathbf{1}} \delta_{\hat{b}_{m \notin \text{s}(\mathbf{i}) \cup \text{s}(\mathbf{j})}, \mathbf{0}}
\end{align}
evaluates to $\pm 1 / 2^{2k - 2l - 1}$ if $\delta_{\hat{b}_{m \in \text{s}(\mathbf{i}) \cap \text{s}(\mathbf{j})}, \mathbf{1}} \delta_{\hat{b}_{m \notin \text{s}(\mathbf{i}) \cup \text{s}(\mathbf{j})}, \mathbf{0}} = 1$ and $|\{ m \in \text{s}(\mathbf{i}) \Delta \text{s}(\mathbf{j}) | U_m = H S^\dagger \}| \equiv 0 \pmod{2}$. Otherwise, it is always zero. This symmetry allows us to further reduce the sample space of $(U_m)$:
\begin{equation*}
o_l = \sum_{\substack{\mathbf{i} < \mathbf{j},\\ |\mathbf{i}| = |\mathbf{j}| = k,\\ |\text{s}(\mathbf{i}) \cap \text{s}(\mathbf{j})| = l}} \sum_{\substack{U_{m \in \text{s}(\mathbf{i}) \Delta \text{s}(\mathbf{j})} \in T \setminus \{\mathbb{I}\},\\ U_{m \notin \text{s}(\mathbf{i}) \Delta \text{s}(\mathbf{j})} = \mathbb{I},\\ |\{ m \in \text{s}(\mathbf{i}) \Delta \text{s}(\mathbf{j}) | U_m = H S^\dagger \}| \equiv 0 \pmod{2}}} \mathbb{E}_{\hat{b}} \left[ V_\text{D} \cdot \binom{n}{k}^{-1} \right].
\end{equation*}
Let $f_l$ denote the uniform distribution over all $\{ \mathbf{i}, \mathbf{j} \}$ such that $|\mathbf{i}| = |\mathbf{j}| = k$ and $|\text{s}(\mathbf{i}) \cap \text{s}(\mathbf{j})| = l$. There are
\begin{equation}
\label{equation:c_l_definition}
c_l \equiv \binom{n}{2k - l} \binom{2k - l}{l} \binom{2k - 2l - 1}{k - l - 1}
\end{equation}
such sets. Then
\begin{align*}
o_l & = \sum_{\substack{\mathbf{i} < \mathbf{j},\\ |\mathbf{i}| = |\mathbf{j}| = k,\\ |\text{s}(\mathbf{i}) \cap \text{s}(\mathbf{j})| = l}} \frac{1}{c_l} \sum_{U_{m \in \text{s}(\mathbf{i}) \Delta \text{s}(\mathbf{j}) - \max (\text{s}(\mathbf{i}) \Delta \text{s}(\mathbf{j}))}^\prime \in T \setminus \{\mathbb{I}\}} \frac{1}{2^{2k - 2l - 1}} \mathbb{E}_{\hat{b}} \left[ \hat{o}_l \left| \begin{aligned} U_{m \in \text{s}(\mathbf{i}) \Delta \text{s}(\mathbf{j}) - \max (\text{s}(\mathbf{i}) \Delta \text{s}(\mathbf{j}))} = U_m^\prime, \\ U_{\max ( \text{s}(\mathbf{i}) \Delta \text{s}(\mathbf{j}) )} = \mathcal{F} (\{ U_m^\prime \}), \\ U_{m \notin \text{s}(\mathbf{i}) \Delta \text{s}(\mathbf{j})} = \mathbb{I} \end{aligned} \right. \right] \\
& = \mathbb{E}_{\{ \mathbf{i}, \mathbf{j} \} \sim f_l, U_{m \in \text{s}(\mathbf{i}) \Delta \text{s}(\mathbf{j}) - \max (\text{s}(\mathbf{i}) \Delta \text{s}(\mathbf{j}))}^\prime \sim g, \hat{b}} \left[ \hat{o}_l \left| \begin{aligned} U_{m \in \text{s}(\mathbf{i}) \Delta \text{s}(\mathbf{j}) - \max (\text{s}(\mathbf{i}) \Delta \text{s}(\mathbf{j}))} = U_m^\prime, \\ U_{\max ( \text{s}(\mathbf{i}) \Delta \text{s}(\mathbf{j}) )} = \mathcal{F} (\{ U_m^\prime \}), \\ U_{m \notin \text{s}(\mathbf{i}) \Delta \text{s}(\mathbf{j})} = \mathbb{I} \end{aligned} \right. \right],
\end{align*}
where $\hat{o}_l \equiv V_\text{D} \cdot \binom{n}{k}^{-1} \cdot c_l \cdot 2^{2k - 2l - 1}$ and $\mathcal{F} (\{ U_m^\prime \}) \in T \setminus \{\mathbb{I}\}$ such that $|\{ U \in \{ U_m^\prime \} \cup \mathcal{F} ( \{ U_m^\prime \} ) | U = H S^\dagger \}| \equiv 0 \pmod{2}$. Therefore, we have $-\binom{n}{k}^{-1} \cdot c_l \leq \hat{o}_l \leq \binom{n}{k}^{-1} \cdot c_l$.

\nocite{*}

\bibliography{apssampv1}

\end{document}